\documentstyle[epsfig,12pt]{article}

\newcommand{\la}[1]{\label{#1}}
\newcommand{\ur}[1]{(\ref{#1})}
\newcommand{\urs}[2]{(\ref{#1},\ref{#2})}

\newcommand{\eq}[1]{eq.~(\ref{#1})}
\newcommand{\eqs}[2]{eqs.~(\ref{#1},\ref{#2})}
\newcommand{\Eq}[1]{Eq.~(\ref{#1})}

\newcommand{\n}{\bf{n}}

\def\Tr{\mbox{Tr}}
\def\beq{\begin{equation}}
\def\eeq{\end{equation}}
\def\bea{\begin{eqnarray}}
\def\eea{\end{eqnarray}}

\begin{document}
\thispagestyle{empty}
\begin{flushright} NORDITA-2000/64 HE
\end{flushright}
\today
\vskip 2true cm
\begin{center}
{\Large\bf On the Non-Abelian Stokes Theorem}
%Reply to Criticism by Faber, Ivanov, Troitskaya and Zach

\vskip 1.5true cm

{\large\bf Dmitri Diakonov$^{\diamond *}$ and Victor Petrov$^*$} \\
\vskip 1true cm
$^\diamond$ {\it NORDITA, Blegdamsvej 17, DK-2100 Copenhagen \O,
Denmark} \\
\vskip .5true cm
$^*$ {\it St.Petersburg Nuclear Physics Institute, Gatchina 188 350,
Russia} \\
\vskip .5true cm

E-mail: diakonov@nordita.dk, victorp@thd.pnpi.spb.ru
\end{center}
\vskip 1.5true cm
\begin{abstract}
\noindent We present the non-Abelian Stokes theorem for the
Wilson loop in various forms and discuss its meaning. Its validity has
been recently questioned by Faber, Ivanov, Troitskaya and Zach.
We demonstrate that all points of their criticism are based on
mistakes in mathematics. Finally, we derive a variant of our formula for
the Wilson loop in lattice regularization.
\end{abstract}

\section{Introduction}

One of the main objects in the Yang--Mills theory is the Wilson
loop or holonomy; it is defined as a path-ordered exponent,

\beq
W_r=
\frac{1}{d(r)}\; \Tr\; {\rm P}\;
\exp\, i\oint\!d\tau\,\frac{dx^\mu}{d\tau}\,A_\mu^a\,T^a,
\la{wl1}\eeq
where $x^\mu(\tau)$ with $0\leq\tau\leq 1$ parametrizes the closed
contour, $A_\mu^a$ is the Yang--Mills field (or connection) and $T^a$
are the generators of the gauge group in a given representation $r$
whose dimension is $d(r)$.

It is generally believed that in three and four dimensions the average
of the Wilson loop in a pure Yang--Mills quantum theory exhibits an
area behavior for large and simple contours (like flat rectangular).
This should be true not for all representations but those
with `$N$-ality' nonequal zero; in the simplest case of the $SU(2)$
gauge group these are representations with half-integer spin $J$.

One of the difficulties in proving the asymptotic area law for the
Wilson loop in half-integer representations (and proving that in
integer representations it is absent) is that the Wilson
loop is a complicated object by itself: it is impossible to calculate it
analytically in a general non-Abelian background field. Meanwhile,
it is sometimes easier to average a quantity over an ensemble than to
calculate it for a specific representative. However, in case of the
Wilson loop the path-ordering is a serious obstacle on that way.

A decade ago we have suggested a formula for the Wilson
loop in a given background belonging to any gauge group and any
representation \cite{DP1}. In this formula the path ordering along the
loop is removed, but at the price of an additional functional integration
over all gauge transformations of the given non-Abelian background field.
This formula is discussed below, in section 2. Furthermore, the
Wilson loop can be presented in a form of a surface integral \cite{DP2},
see section 3. We call this representation the non-Abelian Stokes
theorem. It is quite different from previous interesting statements
\cite{Halp,Ar,Br,Sim} also called by their authors `non-Abelian
Stokes theorems' but which involve surface ordering. Our formula
has no surface ordering. A classification of `non-Abelian Stokes
theorems' for arbitrary groups and their representations has been given
recently by Kondo et al. \cite{Kondo} who used the naturally arising
techniques of flag manifolds.

Though these formulae usually do not facilitate finding Wilson
loops in particular backgrounds, they can be used to average over
ensembles of Yang--Mills configurations, and in more general settings,
see e.g. \cite{DP3,Pol1,Kondo,KAK}.

Our formula for the Wilson loop have been recently questioned by Faber,
Ivanov, Troitskaya and Zach (FITZ) \cite{FITZ}. In section 4 we show
that all points of their criticism are due to mistakes in mathematics,
which we thoroughly locate, one by one.  An alternative formula for the
Wilson loop proposed by FITZ is also mathematically inconsistent, and we
pinpoint their concrete errors.

Finally, in section 5 we present a variant of our formula for the Wilson
loop in lattice regularization. It appears to be very similar to that
presented in section 2.

\section{Formula for the Wilson loop}

Let $\tau$ parametrize the loop defined by the trajectory
$x^\mu(\tau)$ and $A(\tau)$ be the tangent component of the Yang--Mills
field along the loop in the fundamental representation of the gauge
group, $A(\tau)=A_\mu^at^adx^\mu/d\tau$, $\mbox{Tr}(t^at^b) =
\frac{1}{2}\delta^{ab}$. The gauge transformation of $A(\tau)$ is

\beq
A(\tau) \rightarrow S^{-1}(\tau)A(\tau)S(\tau)
+iS^{-1}(\tau)\frac{d}{d\tau}S(\tau).
\la{gt}\eeq
Let $H_i$ be the generators from the Cartan subalgebra ($i=1,...,r;\;
r$ is the rank of the gauge group) and the $r$-dimensional vector
${\bf m}$ be the highest weight of the representation $r$ in which
the Wilson loop is considered. The formula for the Wilson loop derived
in ref. \cite{DP1} is a path integral over all gauge transformations
$S(\tau)$ which should be periodic along the contour:

\beq
W_r=\int DS(\tau)\, \exp \,i\int d\tau\; \Tr\,\left[ m_iH_i\; (S^{-1}AS
+i S^{-1}\dot S)\right].
\la{W1}\eeq
Let us stress that \eq{W1} is manifestly gauge invariant, as is the
Wilson loop itself. For example, in the simple case of the $SU(2)$
group \eq{W1} reads:

\beq
W_J=\int DS(\tau)\, \exp\,i\,J\int d\tau \;\Tr\,\left[\tau_3
(S^\dagger AS +i S^\dagger\dot S)\right]
\la{W12}\eeq
where $\tau_3$ is the third Pauli matrix and
$J=\frac{1}{2},\;1,\; \frac{3}{2},...$
is the `spin' of the representation of the Wilson loop considered.

The path integrals over all gauge rotations \urs{W1}{W12} are not
of the Feynman type: they do not contain terms quadratic in the
derivatives in $\tau$. Therefore, a certain regularization is
understood in these equations, ensuring that $S(\tau)$ is sufficiently
smooth. For example, one can introduce quadratic terms in the angular
velocities $iS^\dagger\dot S$ with small coefficients eventually put to
zero; see ref.\cite{DP1} for details. In ref.\cite{DP1}  \eq{W12} has
been derived in two independent ways: i) by direct discretization and ii)
by using the standard Feynman representation of path integrals as a sum
over all intermediate states, in this case that of an axial top
supplemented by a `Wess--Zumino' type of the action. Another
discretization but leading to the same result has been used recently by
Kondo \cite{Kondo}. A similar formula has been used by Alekseev, Faddeev
and Shatashvili \cite{AFS} who derived a formula for group characters to
which the Wilson loop is reduced in case of a constant $A$ field
actually considered in \cite{AFS}. In ref.\cite{Lun} \eq{W1} has been
rederived in an independent way specifically for the fundamental
representation of the $SU(N)$ gauge group. Finally, at the end of this
paper we consider the Wilson loop in lattice regularization and derive
a formula similar to \eq{W12}.

The second term in the exponent of \eqs{W1}{W12} is in fact a
`Wess--Zumino'-type action, and it can be rewritten not as
a line but as a surface integral inside a closed contour.
Let us consider for simplicity the $SU(2)$ gauge group and parametrize
the $SU(2)$ matrix $S$ from \eq{W12} by Euler's angles,
\bea
\nonumber
S&=&\exp(-i\gamma \tau_3/2)\;\exp(-i\beta\tau_2/2)\;
\exp(-i\alpha\tau_3/2)\\
\nonumber
\\
\la{Euler}
&=&\left(\begin{array}{cc}
\cos\frac{\beta}{2}\,e^{-i\frac{\alpha+\gamma}{2}}&
-\sin\frac{\beta}{2}\,e^{i\frac{\alpha-\gamma}{2}}\\
\sin\frac{\beta}{2}\,e^{-i\frac{\alpha-\gamma}{2}}&
\cos\frac{\beta}{2}\,e^{i\frac{\alpha+\gamma}{2}}
\end{array}\right).
\eea
The derivation of \eq{W12} implies that $S(\tau)$ is a periodic matrix.
It means that $\alpha\pm\gamma$ and $\beta$ are periodic functions of
$\tau$, modulo $4\pi$.

The second term in the exponent of \eq{W12} which we denote by $\Phi$
is then
\bea
\nonumber
\Phi &=& \int d\tau\;\Tr(\tau_3\,i\,S^\dagger\dot S)
=\int d\tau\, \dot \alpha (\cos\beta+\dot \gamma)\\
\la{WZ1}
&=& \int d\tau\, [\dot \alpha (\cos\beta-1)+(\dot \alpha+\dot \gamma)]
=\int d\tau\,\dot \alpha (\cos\beta-1).
\eea
The last term is a full derivative and can be actually dropped because
$\alpha+\gamma$ is $4\pi$-periodic, therefore even for half-integer
representations $J$ it does not contribute to \eq{W12}. Notice that
$\alpha$ can be $2\pi$-periodic if $\gamma$ (which drops from \eq{WZ1})
is $2\pi,\,6\pi,\ldots$-periodic. If $\alpha(1)=\alpha(0)+2\pi\,k$ we
shall say that $\alpha(\tau)$ makes $k$ windings. Integration over all
possible $\alpha(\tau)$ implied in \eq{W12} can be divided into distinct
sectors with different winding numbers $k$.

Let us prove that \eq{WZ1} can be written as an integral over any
surface spanned on the contour (we shall call it a `disk'),

\beq
\Phi=\frac{1}{2}\int d\tau d\sigma\,\epsilon^{abc}\,\epsilon_{ij}
\,n^a\partial_i n^b\partial_j n^c,\qquad i,j=\tau,\sigma,
\la{WZ2}\eeq
where ${\bf n}$ is a unit 3-vector,

\beq
n^a=\frac{1}{2} \Tr\;(S^\dagger\tau^a S \tau_3)
=(\sin\beta\cos\alpha, \,\sin\beta\sin\alpha, \,\cos\beta).
\la{n}\eeq
It is understood that ${\bf n}$ or $\alpha$ and $\beta$ are continued
inside the disk. We denote the second coordinate by $\sigma$ such
that $\sigma=1$ corresponds to the edge of the disk coinciding with the
contour and $\sigma=0$ corresponds to the center of the disk.
Let us establish what conditions should the continuation
${\bf n}(\sigma,\tau)$ satisfy.

\Eq{WZ2} can be identically rewritten as
\bea
\nonumber
\Phi&=&\int\!\int\! d\tau
d\sigma\left[\partial_\tau\alpha\partial_\sigma
(\cos\beta-1)-\partial_\sigma\alpha\partial_\tau(\cos\beta-1)\right]\\
\nonumber
&=&\int\!\int\! d\tau
d\sigma\left\{\partial_\sigma\left[\partial_\tau\alpha(\cos\beta-1)\right]
-\partial_\tau\left[\partial_\sigma\alpha(\cos\beta-1)\right]\right\}\\
\la{WZ21}
&=&\int_0^1\!d\tau
\left[\partial_\tau\alpha\,(\cos\beta-1)\right]^{\sigma=1}_{\sigma=0}
-\int_0^1\!d\sigma
\left[\partial_\sigma\alpha\,(\cos\beta-1)\right]^{\tau=1}_{\tau=0}.
\eea
The second term here is zero, for the following reasons. Let
$\alpha(\tau)$ belong to the sector with winding number $k$. The
periodicity property of $\alpha(\sigma,\tau)$ cannot change abruptly as
we continue it inside the disk. We can write $\alpha(\sigma,\tau)=2\pi
k\tau +\tilde\alpha(\sigma,\tau)$ where $\tilde\alpha$ is strictly
periodic, as well as $\cos\beta$. Therefore, the integrand of the
second term in \eq{WZ21} is zero by periodicity.

Let us now consider the first term in \eq{WZ21}. We want the surface
integral \ur{WZ2} to reproduce the contour integral \ur{WZ1}, up to a
possible contribution of $4\pi l$ with integer $l$, coming from the
center of the disk, $\sigma=0$. Such a contribution is allowed since it
does not affect \eq{W12} even for half-integer $J$. The contribution of
the first term in \eq{WZ21} at $\sigma=0$ is

\beq
-\int_0^1 d\tau\,[2\pi k+\partial_\tau\tilde\alpha(0,\tau)]
[\cos\beta(0,\tau)-1]=2\pi k[\cos\beta(0)-1]=4\pi l,
\la{sigma0}\eeq
where we have used that $\tilde\alpha$ is periodic while
$\beta(0,\tau)$ is in fact independent of $\tau$, otherwise
$n^3=\cos\beta$ would be not uniquely defined at the center of the
disk. \Eq{sigma0} means that either $\alpha(\tau)$ belongs
to the winding number $k=0$ sector, then $\beta(0)$ is arbitrary,
or, if $k\neq 0$ then $\beta=0,\pi$ meaning that $n^3=\pm 1$ at the
center of the disk. Notice that for $|n^3|\neq 1$ the components
$n^{1,2}$ are nonzero and are varying very rapidly as function of
$\tau$ at the point $\sigma=0$. These conditions can be summarized as
the condition that ${\bf n}(\sigma,\tau)$ should be smooth and uniquely
defined inside the disk.

Let us note that if the surface is closed or infinite the r.h.s. of
\eq{WZ2} is the integer topological charge of the ${\bf n}$ field on
the surface:

\beq
Q=\frac{1}{8\pi}\int d\sigma d\tau\,\epsilon^{abc}\,\epsilon_{ij}
\,n^a\partial_i n^b\partial_j n^c.
\la{topc}\eeq

\Eq{WZ2} can be also rewritten in a form which is invariant under
the reparametrizations of the surface. Introducing the invariant
element of a surface,

\beq
d^2S^{\mu\nu}=d\sigma\,d\tau\;\left(
\frac{\partial x^\mu}{\partial \tau}
\frac{\partial x^\nu}{\partial \sigma}-
\frac{\partial x^\nu}{\partial \tau}
\frac{\partial x^\mu}{\partial \sigma}\right)
= \epsilon^{\mu\nu}\;d({\rm Area}),
\la{elsur}\eeq
one can rewrite \eq{WZ2} as

\beq
\Phi=\frac{1}{2}\int\; d^2S^{\mu\nu}
\epsilon^{abc} n^a\partial_\mu n^b\partial_\nu n^c.
\la{WZ3}\eeq
We get for the Wilson loop

\beq
W_J=\int D{\bf n}(\sigma,\tau)\;\exp\left[iJ\int\,d\tau
(A^an^a)+\frac{iJ}{2}\int\;d^2S^{\mu\nu}
\epsilon^{abc} n^a\partial_\mu n^b\partial_\nu n^c\right].
\la{W2}\eeq

The interpretation of this formula is obvious: the unit vector $\n$ plays
the role of the instant direction of the color `spin' in color space;
however, multiplying its length by $J$ does not yet guarantee that we
deal with a true quantum state from a representation labelled by $J$ --
that is achieved only by introducing the `Wess--Zumino' term in \eq{W2}:
it fixes the representation to which the probe quark of the Wilson loop
belongs to be exactly $J$.

\section{Non-Abelian Stokes theorem}

We can now rewrite the exponent in \eq{W2} so that both terms
appear to be surface integrals \cite{DP2}:

\beq
W_J=\int D{\bf n}(\sigma,\tau)\;\exp\frac{iJ}{2}\int
d^2S^{\mu\nu}\left(-F_{\mu\nu}^an^a+
\epsilon^{abc}\, n^a\left(D_\mu n\right)^b\left(D_\nu
n\right)^c\right),
\la{W3}\eeq
where
$D_\mu^{ab}=\partial_\mu\delta^{ab}+\epsilon^{acb}A_\mu^c$
is the covariant derivative and $F_{\mu\nu}^a = \partial_\mu~
A_\nu^a - \partial_\nu~A_\mu^a + \epsilon^{abc}~A_\mu^b~A_\nu^c$
is the field strength. Indeed, expanding the exponent of \eq{W3} in
powers of $A_\mu$ one observes that the quadratic term cancels out
while the linear one is a full derivative reproducing the $A^an^a$ term
in \eq{W2}; the zero-order term is the `Wess--Zumino' term \ur{WZ2} or
\ur{WZ1}. Note that both terms in \eq{W3} are explicitly gauge
invariant. We call \eq{W3} the non-Abelian Stokes theorem. We stress
that it is different from previously suggested Stokes-like
representations of the Wilson loop, based on ordering of elementary
surfaces inside the loop \cite{Halp,Ar,Br,Sim}. It is understood that
the functional integration measure in \eq{W3} is such that $W_J=1$
for zero field.

The gauge-invariant field strength appearing in \eq{W3},

\beq
G_{\mu\nu}=F_{\mu\nu}^an^a
- \epsilon^{abc} n^a\left(D_\mu n\right)^b\left(D_\nu n\right)^c,
\la{gifs}\eeq
coincides in form with the field strength introduced by Polyakov
\cite{Pol2} and 't Hooft \cite{tH} in connection with monopoles. In
that case the unit-vector field $n^a$ has the meaning of the
direction of the elementary Higgs field, $\phi^a/|\phi|$.

One can introduce a `monopole current',

\beq
j_\mu^* = \frac{1}{8\pi}\epsilon_{\mu\nu\kappa\lambda}\partial_\nu
G_{\kappa\lambda}.
\la{moncur}\eeq
Using equations
\bea
\nonumber
\epsilon_{\mu\nu\kappa\lambda}D_\nu^{ab} F_{\kappa\lambda}^b&=&0,
\qquad
\epsilon_{\mu\nu\kappa\lambda}\epsilon^{abc}(D_\nu n)^a(D_\kappa n)^b
(D_\lambda n)^c = 0, \\
\nonumber
\epsilon_{\mu\nu\kappa\lambda}(D_\nu D_\kappa n)&=&0,\qquad
n^a(D_\kappa n)^a =0,
\eea
one can easily check that this current is zero. In mathematical language
it is the statement that any exact two-form is complete. There is a
subtlety here, however. Let us consider a configuration which we shall
call the `Wu--Yang monopole', $n^a=x^a/r$ (not to be confused with the
monopoles of the $A_\mu$ field). Let us take a contour of unit radius
about the origin lying in the $z=0$ plane, the equator. In terms of
angles the monopole corresponds to $\alpha(\tau)=\phi=2\pi\tau$,
$\beta=\theta$, see \eq{n}. At the equator $\theta=\pi/2$, and we have
from the line form \ur{WZ1}:

\beq
\Phi=\int_0^1 d\tau\,\partial_\tau\alpha\,(\cos\beta-1)
=\int_0^1 d\tau\, 2\pi\,(-1)=-2\pi.
\la{lf}\eeq
Let us now compute $\Phi$ in the surface form: first, over the upper
hemisphere whose edge is the equator, second, over the lower hemisphere.
For the upper hemisphere we introduce the variable
$\sigma=1-\cos\theta$, so that $\sigma=0$ at the north pole and
$\sigma=1$ on the equator. Then $\cos\beta-1=-\sigma$,
and \eq{WZ2} reads

\beq
\Phi=\int_0^1\int_0^1d\tau d\sigma\,\partial_\tau(2\pi\tau)
\partial_\sigma(-\sigma)=-2\pi,
\la{sf1}\eeq
in agreement with \eq{lf}. For the lower hemisphere we define the
variable $\sigma=1+\cos\theta$ so that $\sigma=0$ at the south pole and
$\sigma=1$ on the equator. Then $\cos\beta-1=\sigma-2$,
and \eq{WZ2} reads

\beq
\Phi=\int_0^1\int_0^1d\tau d\sigma\,\partial_\tau(2\pi\tau)
\partial_\sigma(\sigma-2)=+2\pi.
\la{sf2}\eeq
There is no arithmetical contradiction here since in deriving the
surface form \ur{WZ21} we have allowed for the contribution from
$\sigma=0$ to be a multiple of $4\pi$, and it is exactly what has
happened here. Simultaneously, it is in correspondence with the
fact that the integral over the full sphere around a monopole (see
\eq{topc}) is $4\pi$. In both cases the contribution of the `Wu--Yang
monopole' configuration to the functional integral \ur{W3} is

\beq
\exp\,\pm\,2\pi\,i\,J=(-1)^{2J},
\la{moncontr}\eeq
irrespective of whether the surface is drawn above or below the
singularity of the ${\bf n}$ field. Notice that \eq{moncontr}
makes a clear distinction between integer and half-integer
representations.

If, however, we take the surface which passes exactly through the center
of the monopole, we shall be in trouble. For example, choosing the
equatorial plane we have $n^3=0$ everywhere on the plane, hence the
surface integral \ur{WZ2} is zero, in contradiction with \eq{lf}. It is
because we have drawn the surface directly through the singularity and
have thus violated the condition needed for the applicability of the
surface form \ur{WZ2}. Does it mean that singular configurations of the
type $n^a=x^a/r$ are altogether forbidden in the functional integral
\ur{W3}? Probably not, as long as such singularities have a
zero-measure probability to hit the surface.

Let us integrate the `monopole charge density' $j_0^*$ over a
volume surrounded by a closed surface. According to the Gauss theorem,

\beq
8\pi\int d^3V_\mu j_\mu^*=-\int d^2S^{\mu\nu}\epsilon^{abc}
n^a\partial_\mu n^b\partial_\nu n^c =-8\pi\,Q,
\la{moncharge}\eeq
where $Q$ is the integer winding number \ur{topc} of the $\n$ field over
the closed 2-surface surrounding the `monopole'; in the concrete
case of the field $n^a=x^a/r$ one has $Q=1$. Since the surface can be
drawn as close to the center as one wishes, it is tempting to say that
the `Wu--Yang monopole' has $j_0^*=-\delta^{(3)}(r)$. Since computing
$j_0^*$ involves singular operations, one can regularize the
singularity, for example by replacing
$n^a\to x^a/(r^2+\delta^2)^{1/2},\;\delta\to 0$.
This regularization, indeed, leads to
$j_0^*=-3\delta^2/(r^2+\delta^2)^{5/2}/(4\pi)\to -\delta^{(3)}(r)$.
It should be stressed, however, that this regularization violates the
condition ${\bf n}^2=1$ which has been crucial to prove that
$j_\mu^*=0$.

To conclude this discussion: The Wilson loop is defined as a
contour integral, so when one writes it in a surface form one has
to take care that it does reproduce the contour form. Generally,
singular ${\bf n}(\sigma,\tau)$ leading, after regularization,
to nonzero $j^*_\mu$ imply that the result depends on how one draws the
surface, which is unacceptable. However, e.g. a `gas' of singularities
of the type $n^a=x^a/r$ with quantized $4\pi$ flux (in $d=4$ it is a
`gas' of worldlines) is still admissible since the probability that
the singularity occurs exactly on the surface has zero measure.

Let us briefly discuss gauge groups higher than $SU(2)$: for that
purpose we have to return to our \eq{W1}. \Eq{W1} is valid
for any group and any representation. It is easy to present it also in a
surface form. We denote the combination of Cartan generators $m_iH_i$
where ${\bf m}$ is the highest weight of a given representation $r$ by
${\cal H}_r$. Using the identity
\bea
\nonumber
\epsilon_{ij}\,\partial_i\,\Tr\,{\cal H}_rS^{-1}\nabla_{\!j}S&=&
\epsilon_{ij}\,\Tr\,{\cal H}_r\left[S^{-1}\left(\nabla_{\!i}
\nabla_{\!j}S\right)
+\left(S^{-1}\overleftarrow\nabla_{\!i}\right)\left(\nabla_{\!j}
S\right)\right]\\
\la{stokes1}
&=&\epsilon_{ij}\,\Tr\,{\cal H}_r\left[-\frac{i}{2}(S^{-1} F_{ij}S)
+\left(S^{-1}\overleftarrow\nabla_{\!i}
\right)\left(\nabla_{\!j} S\right)\right],\\
\nonumber
\nabla_{\!i}&=&\partial_i-iA^a_it^a,
\qquad \overleftarrow\nabla_{\!i}=\overleftarrow\partial_i+iA^a_it^a,
\eea
we can present \eq{W1} in a surface form:

\beq
W_r=\int DS(\sigma,\tau)\,\exp\,i\int dS^{\mu\nu}\,
\Tr\,{\cal H}_r\left[-\frac{i}{2}(S^{-1} F_{\mu\nu}S)
+\left(S^{-1}\overleftarrow\nabla_{\!\mu} \right)\left(\nabla_{\!\nu}
S\right)\right].
\la{W14}\eeq

Actually, \eqs{W1}{W14} depend not on all parameters of the
gauge transformation but only on those which do not commute with the
Cartan combination ${\cal H}_r=m_iH_i$. In the $SU(2)$ case one
has $m_iH_i= J\tau_3,\;J=1/2,1,3/2,\ldots$, since $SU(2)$ is of rank 1,
and there is only one Cartan generator. Therefore, in the $SU(2)$ case
one integrates over the coset $SU(2)/U(1)$ for any representation;
this coset can be parametrized by the ${\bf n}$ field as described
above.

In case of higher groups the particular coset depends on the
representation of the Wilson loop. For example, in case the Wilson loop
is considered in the fundamental representation of the $SU(N)$ group the
combination $m_iH_i$ is proportional to one particular generator of the
Cartan subalgebra, which commutes with the $SU(N-1)\times U(1)$ subgroup.
[In case of $SU(3)$ this generator is the Gell-Mann $\lambda_8$
matrix or a permutation of its elements.] Therefore, the appropriate
coset for the fundamental representation of the $SU(N)$ group is
$SU(N)~/SU(N-1)~/U(1)= CP^{N-1}$. A possible parametrization of this
coset is given by a complex $N$-vector $u^\alpha$ of unit length,
$u_\alpha^\dagger u^\alpha=1$. To be concrete, the Cartan combination
in the fundamental representation can be always set to be
$m_iH_i={\rm diag}(1,0,\ldots,0)$ by rotating the axes and
subtracting the unit matrix. In such a basis $u^\alpha$ is just the
first column of the unitary matrix $S$ while $u_\alpha^\dagger$
is the first row of $S^\dagger$. Unitarity of $S$ implies that
$u_\alpha^\dagger u^\alpha=1$.

In this parametrization \eq{W1} can be written as

\beq
W^{SU(N)}_{{\rm fund}}=\int Du\,Du^\dagger\,
\delta(u_\alpha^\dagger u^\alpha-1)\,
\exp i\!\int\! d\tau\,\frac{dx^\mu}{d\tau}
u^\dagger_\alpha\left(i\nabla_\mu\right)^\alpha_\beta\,u^\beta.
\la{W13}\eeq
Using the identity,
\bea
\nonumber
\epsilon_{ij}\,\partial_i\left(u^\dagger\nabla_ju\right)&=&
\epsilon_{ij}\,\left[\left(\nabla_i u\right)^\dagger
\left(\nabla_j u\right)+u^\dagger\nabla_i\nabla_ju\right]\\
\la{stokes2}
&=&\epsilon_{ij}\,\left[-\frac{i}{2}(u^\dagger F_{ij}u)+
\left(\nabla_i u\right)^\dagger\left(\nabla_j u\right)\right],
\eea
we can present \eq{W13} in a surface form:

\beq
W^{SU(N)}_{{\rm fund}}=\int Du\,Du^\dagger\,
\delta(|u|^2-1)\,
\exp i\!\int\! dS^{\mu\nu}
\left[\frac{1}{2}(u^\dagger F_{\mu\nu}u)+i
\left(\nabla_\mu u\right)^\dagger\left(\nabla_\nu u\right)\right],
\la{W15}\eeq
where $F_{\mu\nu}$ is the field strength in the fundamental
representation. \Eq{W14} has been first published in ref.\cite{Lun}
however with an unexpected overall coefficient 2 in the exponent.
\Eq{W15} presents the non-Abelian Stokes theorem for the Wilson loop
in the fundamental representation of $SU(N)$. In the particular case
of the $SU(2)$ group transition to \eq{W3} is achieved by identifying
the unit 3-vector: $n^a=u^\dagger_\alpha(\tau^a)^\alpha_\beta u^\beta$
where

\beq
u^\alpha=\left(\begin{array}{c}\cos\frac{\beta}{2}\,
e^{-i\frac{\alpha+\gamma}{2}}\\
\sin\frac{\beta}{2}\,
e^{\,i\frac{\alpha-\gamma}{2}}\end{array}\right),\qquad
2i\,u^\dagger\partial_\tau u
=\dot\alpha(\cos\beta-1)+(\dot\alpha+\dot\gamma).
\la{u}\eeq

It should be mentioned that the quantity

\beq
\int d\sigma d\tau\, \epsilon_{ij}\,i\,\partial_iu^\dagger_\alpha
\partial_ju^\alpha= 2\pi Q
\la{topzar}\eeq
appearing in \eq{W15} is the topological charge of the 2-dimensional
$CP^{N-1}$ model. For closed or infinite surfaces $Q$ is an integer.
For example, in case of the `Wu--Yang monopole' $u^\alpha$ can be
chosen in two forms compatible with periodicity:

\beq
u^\alpha=\left(\begin{array}{c}\cos\frac{\theta}{2}\\
\sin\frac{\theta}{2}\,
e^{i\phi}\end{array}\right)\qquad{\rm or}\qquad
\left(\begin{array}{c}\cos\frac{\theta}{2}\,e^{-i\phi}\\
\sin\frac{\theta}{2}\end{array}\right).
\la{umon}\eeq
They are regular everywhere except the negative (positive) $z$ axis.
One can regularize the string singularity of $u^\alpha$ relaxing the
condition $|u|^2=1$. The coefficient $2\pi$ guarantees that \eq{W15}
does not actually depend on the choice of the surface even for singular
$u$'s, cf. the discussion around \eq{moncontr} above. At the same
time it means that $CP^{N-1}$ `instantons' of an effective
2-dimensional model, having integer topological charges, can hardly be
relevant to the area behavior of the Wilson loop, as conjectured
recently in ref.\cite{Kondo}. The coefficient with which the
topological charge enters the formula for the Wilson loop \ur{W15}
corresponds to the `instanton angle' $\theta=2\pi$, hence it is
unobservable from the point of view of the 2-dimensional instantons.
Only configurations with half-integer topological charges (like that
given by \eq{umon} or its gauge equivalents) stand a chance of being of
relevance to confinement.

In case the Wilson loop is taken in the adjoint representation of
the $SU(N)$ gauge group the combination $m_iH_i$ in \eq{W1} is the
highest root. Only group elements of the form $\exp(i\alpha_iH_i)$
commute with this combination, belonging to the maximal torus
subgroup $U(1)^{N-1}$. Hence, in case of the adjoint representation
one in fact integrates over the maximal coset
$SU(N)/U(1)^{N-1}=F^{N-1}$, i.e. over flag variables \cite{Per,Kondo}.

\section{Reply to criticizm by Faber, Ivanov, Troitskaya and Zach}

In a recent paper \cite{FITZ} Faber, Ivanov, Troitskaya and Zach
(FITZ) have questioned the validity of our formula for the Wilson
loop, \eq{W12}. Their points can be summarized as follows: \\

\noindent 1. A direct calculation using \eq{W12} in two simple
cases where the Wilson loop is known, i.e. for a pure gauge potential
and for a vortex field, produces zero results instead of correct
ones. \\

\noindent 2. The regularized evolution operator for an axial top used
by us to derive the formula for the Wilson loop, when computed
directly, is zero. \\

\noindent 3. The evaluation of this evolution operator by introducing
small regulator moments of inertia, as suggested by us, is prohibited
because it changes the symmetry from $SU(2)$ to that of $U(2)$.  \\

\noindent 4. Another formula for the Wilson loop is proposed which is
similar to but different from our.\\

We consider these points below, one by one, and show that all four
are groundless as due to errors in mathematics.\\

{\bf 1.} In sections 4,5 of their paper FITZ attempt to compute the
Wilson loop in two simple cases: i) in a pure gauge background and ii)
in a background of a vortex. In both cases the Wilson loop is known.
In this calculation FITZ use our discretization of the path integral
though without the regulator terms which, as we have stressed in the
original paper \cite{DP1}, are important to get the correct result.
Nevertheless, it could be a useful exercise, were it not for mistakes
in mathematics.

In this calculation FITZ use several relations for group characters
citing a paper by Bars \cite{Bars} ``modified for our case". The
modification is not performed properly. Ref. \cite{Bars} deals with
the $GL(N)$ group, and averaging is performed over the $U(N)$ group,
not $SU(N)$ which is the case in question. The characters of a
group are unambiguously defined for the elements of that group.
Meanwhile, FITZ intensively use such ambiguous quantities as $SU(2)$
characters of the non-$SU(2)$ elements like $t_3,\;t_3^2=1/4$,  etc.
($t_3$ is one half of the Pauli matrix $\tau_3$). Though no explicit
definition of the characters of non-$SU(2)$ elements is given
in the paper, from eqs. (64) and (A3) of ref.\cite{FITZ} one
infers that FITZ implicitly use the definition:

\beq
\chi_j[(t_3)^n\,U]=\sum_{m=-j}^jm^nD^j_{mm}(U)
\la{char1}\eeq
where $D^j_{mn}(U)$ are Wigner's finite rotation matrices,
$\sum_m D^j_{mm}(U)=\chi_j[U]$. [The properties of $D$-functions
are given, e.g., in ref. \cite{MVK}.]

The key formula of FITZ calculation is their decomposition formula
(51),

\beq
\exp\,z\Tr[t_3\,U]=\sum_j a_j(z)\, \chi_j[t_3\, U]
=\sum_j a_j(z)\, \sum_{m=-j}^j m\,D^j_{mm}(U),
\la{wrong1}\eeq
and its inverse (eq. (53) of \cite{FITZ}),

\beq
a_j(z)=\frac{3}{j(j+1)}\int dU\chi_j[t_3\,U^\dagger]\,
\exp \left(z\Tr[t_3\,U]\right).
\la{wrong2}\eeq
The quantities $\chi_j[t_3\,U]$ do not form a complete set of
functions for the space of $SU(2)$ matrices, therefore there are no
reasons why the decomposition \ur{wrong1} is at all possible, for
whatever coefficients $a_j(z)$.  To see that it is in fact wrong let us
present the decomposition of the l.h.s.  of \eq{wrong1} using
well-defined characters of group elements only. We make use of the fact
that $t_3=(-i/2)(i\tau_3)=(-i/2)\exp(i\pi t_3)$ where the last factor
is definitely an element of $SU(2)$.

We have
\[
\exp\,z\Tr[t_3\,U]=\exp\left\{(-iz/2)\Tr[e^{i\pi t_3}\,U]\right\}=
\sum_j \tilde a_j(z) \chi_j\left[e^{i\pi t_3}\,U\right]
\]
\beq
= \sum_j \tilde a_j(z)\, \sum_{m=-j}^je^{i\pi m}\,D^j_{mm}(U),
\qquad \tilde a_j(z)=e^{-i\pi j}\,(2j+1)\,\frac{2J_{2j+1}(z)}{z},
\la{corr1}\eeq
the coefficients $\tilde a_j(z)$ being well-known from the lattice
strong-coupling expansion \cite{DZ}.

\Eq{corr1} differs significantly from \eq{wrong1} suggested by FITZ;
there exists no choice of coefficients $a_j(z)$ for which \eq{wrong1}
coincides with the correct one, \eq{corr1}. \Eq{wrong1} and its inverse,
\eq{wrong2}, (eqs. (51) and (53) of \cite{FITZ}) result from
uncritical `modification' of ref.\cite{Bars} and are wrong.

Furthermore, assuming the decomposition \ur{wrong1} to hold true
FITZ arrive to a  ``completeness condition'' for their coefficients
$a_j(z)$ (see their eq.(59)):

\beq
a_0^2(z)+\sum_{j>0}\frac{1}{3}j(j+1)a_j^2(z)=1,
\la{compl1}\eeq
where (see eq.(A6) of \cite{FITZ})
\beq
a_j(z) = \left\{\begin{array}{cc}
\frac{2}{z}J_1(z), &j=0,\\
\frac{3(2j+1)}{j(j+1)}\,\frac{J_{2j+1}(z)}{z}, & j =
1/2,3/2,5/2,\ldots,\\
0, & j = 1,2,\ldots
\end{array}\right.\,.
\la{A6}\eeq

%%%%%%%%%%
% FIGURE 1
%%%%%%%%%%
\begin{figure}
\centerline{\epsfxsize10.0cm\epsffile{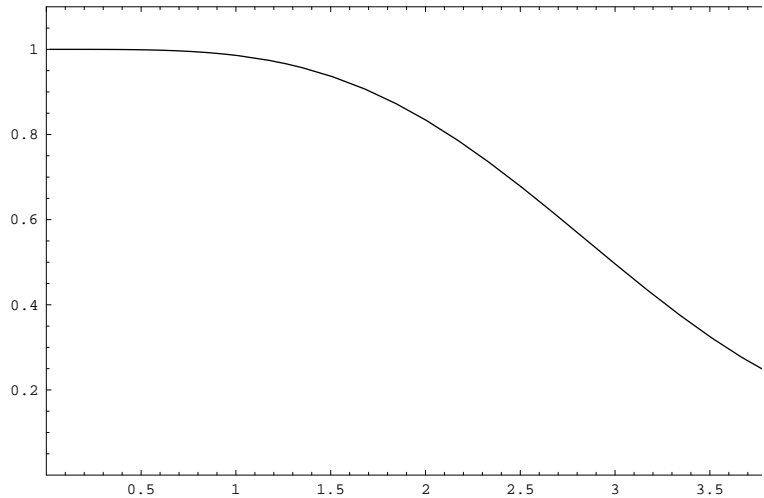}}
\caption[]{The l.h.s. of the ``completeness condition''
claimed by FITZ to be identically unity.}
\end{figure}

We plot the l.h.s. of \ur{compl1} as function of $z$ in Fig.1: it does
not look as being identically unity, as claimed by FITZ. It is another
manifestation of that \eqs{wrong1}{wrong2} are wrong
\footnote
{Ironically, FITZ have checked their ``completeness condition'' themselves
but only at one value of $z=1$ where the l.h.s. of \eq{compl1} is still
rather close to unity being equal to 0.986. Having computed this value
summing up the series on the l.h.s. up to $j=\frac{5}{2}$ FITZ note:
``Thus, the series converges slowly to unity'' not paying attention
to that their $j=\frac{3}{2}$ contribution is $\sim 10^{-4}$ and the
next contribution at $j=\frac{5}{2}$ is $\sim 10^{-9}$. The series
converges rather rapidly but not to unity.}.

Unfortunately, this ``completeness condition'' together with the
erroneous coefficients $a_j(z)$ are at the heart of the calculation
of the Wilson loop by FITZ both for the pure gauge and vortex cases,
which cannot be, thus, considered as correct. \\

{\bf 2.} In section 6 of their paper FITZ attempt to compute the
(regularized) evolution operator for the `Wess--Zumino' action,
following directly our approach. This calculation has been presented in
some detail in the original paper \cite{DP1}, however, FITZ seem
to be dissatisfied by it and present their own. Their final answer
(eqs. (98) and (112) of ref. \cite{FITZ}), which differs from our,
is a result of several mistakes.

First, going from eq.(83) to eq.(87) FITZ use a strange relation,
\beq
\exp\left(\sum_{n=0}^N(-i)\frac{I_\perp}{2\delta}(-4)\right)
=\exp\left(iN(N+1)\frac{I_\perp}{\delta}\right),
\la{osh1}\eeq
instead of the correct (and trivial) $\exp(i2(N+1)I_\perp/\delta)$,
where $I_\perp$ and $\delta$ are constants and $N$ is the number
of pieces in which one divides the contour.

Second, and more important, both equations in (91) are erroneous, they
do not follow from eq.(89) from where they are derived. Passing from
eq.(89) to eq.(91) one gets:
\bea
\Tr(R_nR_{n+1}^\dagger)&=&
2-\frac{1}{4}\left[\delta\alpha_n^2+\delta\beta_n^2
+\delta\gamma_n^2+2\delta\alpha_n\delta\gamma_n\cos\beta_n\right],\\
\nonumber
\Tr(R_nR_{n+1}^\dagger\tau_3)&=&
i(\delta\alpha_n+\delta\gamma_n\cos\beta_n),\\
\nonumber
\delta\alpha_n&=&\alpha_{n+1}-\alpha_n,\quad
\delta\beta_n=\beta_{n+1}-\beta_n,\quad
\delta\gamma_n=\gamma_{n+1}-\gamma_n.
\la{osh2}\eea
FITZ have written these formulae without the crucial factors
$\cos\beta_n$. Because of this mistake the subsequent integration
over Euler angles $\alpha,\beta,\gamma$ becomes gaussian,
and the evolution operator for the axial top, as computed by FITZ,
in fact becomes that of a free particle, which is definitely wrong.
The factors $\cos\beta_n$ being reinstalled, the derivation returns to
that of our paper \cite{DP1}.\\

{\bf 3.} The last objection by FITZ is to our alternative (and
in fact equivalent) derivation of the evolution operator, this time
through the standard Feynman representation for the path integral as a
sum over intermediate states. FITZ quote our result for the evolution
operator of an axial top with the `Wess--Zumino' term, evolving from
its orientation given by a unitary matrix $R_1$ at time $t_1$ to
orientation $R_2$ at time $t_2$:

\beq
{\cal Z}_{{\rm Reg}}(R_2,R_1)=(2J+1) D^J_{JJ}(R_2R_1^\dagger)\,
\exp\left[-i(t_2-t_1)\frac{J}{2I_\perp}\right]
\la{nash}\eeq
where $I_\perp$ is a regulator moment of inertia, $I_\perp\to 0$.
Apart from a nontrivial dependence on the orientation matrices
$R_{1,2}$ coming through the Wigner $D$-function, this expression
contains a phase factor $\exp(-i(t_2-t_1)...)$. It is
an overall factor independent of the external field:
it can and should be absorbed into the integration measure to make the
evolution operator unity for the trivial case $R_2=R_1=1$.
Indeed, dividing the time interval into $N$ pieces of small
length $\delta,\;N\delta=t_2-t_1$, one can write this factor as
a product,

\beq
\exp\left[-i(t_2-t_1)\frac{J}{2I_\perp}\right]
=\prod_{k=1}^N\exp\left[-i\delta \frac{J}{2I_\perp}\right],
\la{factor1}\eeq
where, according to the regularization prescription of ref. \cite{DP1},
$\delta/I_\perp\ll 1$ so that each factor is close to unity. Each
factor can be now absorbed into the integration measure $dR(t_k)$ in the
functional-integral representation for the evolution operator
\ur{nash}. The fact that the factor is complex is irrelevant; moreover,
it is typical for the path-integral representation of the evolution
operators to have a complex measure, see the classical Feynman's book
\cite{F}.  However, FITZ write:  ``...a removal of the fluctuating
factor is prohibited since this leads to the change of the starting
symmetry of the system from $SU(2)$ to $U(2)$''.  It may seem that FITZ
believe that an absorption of a constant factor into the integration
measure changes the number of variables over which one integrates.\\

{\bf 4.} FITZ make an attempt to derive another formula for the Wilson loop
(see section 2 of ref. \cite{FITZ}); this derivation is, however,
inconsistent. Their resulting path-integral representation for the
Wilson loop is (see eq.(30) of \cite{FITZ}; we simplify their
notations to make the formulae better readable):

\beq
W_J=\frac{1}{(2J+1)^2}\int \!D\Omega(\tau)
\sum_{\{m(\tau)\}}(2J+1)\,\exp\left(i\oint_{C}d\tau \,m(\tau)\,
\sqrt{2\Tr\left[A^\Omega(\tau) A^\Omega(\tau)\right]}\right),
\la{30}\eeq
where $A^\Omega=\Omega A\Omega^\dagger-i\dot\Omega\Omega^\dagger$
is the gauge-transformed Yang--Mills field tangent to the
contour parametrized by $\tau,\quad 0\leq \tau\leq 1,$ such that
$A=A_\mu\, dx^\mu/d\tau,\quad \dot\Omega=\partial_\mu\Omega\,
dx^\mu/d\tau$.  One integrates over all gauge transformations
$\Omega(\tau)$ in a given representation of the $SU(2)$ group,
labelled by spin $J$, and sums over the projections $m$ from
$-J$ to $+J$ {\em at all points of the loop}.

Integration over a variable $m(\tau)$ assuming only integer or
half-integer values is a unusual construction and one can
question whether such an integral has a limiting value as one takes
the number of discretization points $N$ to infinity. We have
checked that \eq{30} {\em does not possess a finite limiting value} at
least for the zero field, $A=0$. However, this is not our main point.
The main problem is the derivation of \eq{30} by FITZ.

If one looks into how \eq{30} has been derived by FITZ one finds
that it is a result of two mistakes in mathematics.

FITZ start from using the following discretized form of the
Wilson loop (see their eqs. 14,15):

\[
W_r=\frac{1}{d_r^2}\lim_{n\to\infty}\int\ldots\int
d\Omega(\tau_1)\ldots d\Omega(\tau_N)
\]
\beq
d_r\chi_r\left[\Omega(\tau_N)(1+iA\, \Delta \tau)
\Omega^\dagger(\tau_{N-1})\right]\ldots
d_r\chi_r\left[\Omega(\tau_1)(1+iA\,\Delta \tau)
\Omega^\dagger(\tau_N)\right],\qquad \Delta\tau=\tau_k-\tau_{k-1},
\la{vern}\eeq
where gauge transformations $\Omega$ and the gauge field
$A$ are taken in a given representation $r$ with dimension $d_r$;
$\chi_r$ is the group character.
%We have nothing against this presentation of the Wilson loop.
Indeed, integrating over gauge transformations at all points
along the loop $\Omega(\tau_k)$ and using the relation

\beq
\int d\Omega(\Omega^{\dagger})_{a_1b_1}(\Omega)_{a_2b_2}
= \frac{1}{d_r}\,\delta_{a_1b_2}\,\delta_{b_1a_2},
\la{compl2}\eeq
one recovers the Wilson loop as a path-ordered product of
the factors $(1+iA\Delta\tau)$ along the loop. However at this point
FITZ depart from a consistent derivation.

It is tempting to say that, since the discretization points on the
contour can be taken as close to one another as one wishes, the gauge
transformations $\Omega$ at neighboring points are also close. It
should be emphasized that this is, generally, wrong: one needs a
{\em fully independent} integration over $\Omega$'s at
neighbor points -- otherwise the unity matrix on the r.h.s. of
\eq{compl2} is not achieved. In other words the unitary matrix
$\Omega(\tau_k)\Omega^\dagger(\tau_{k-1})$ is, generally, not close to
the unity matrix, even though the points $\tau_k$ and $\tau_{k-1}$ are
close. The derivative $\dot\Omega$ does not exist in a strict
sense since taking the neighbor points closer does not change the fact
that $\Omega(\tau_k)$ and $\Omega(\tau_{k-1})$ are independent
integration variables.

This important circumstance is neglected by FITZ who say: ``Due to
the infinitesimality of the segments we can omit the path ordering
operator" and write (see their eq. (18))

\beq
\Omega(\tau_k)(1+iA\Delta\tau)\Omega^\dagger(\tau_{k-1})
= \exp \,i\int_{\tau_{k-1}}^{\tau_k} d\tau A^\Omega(\tau).
\la{18}\eeq
This equation is erroneous for reasons explained above. Only if
$\Omega(\tau)$ has a finite derivative one can expand
$\Omega(\tau_k)\Omega^\dagger(\tau_{k-1})=
1+\dot\Omega(\tau_{k-1})\Omega^\dagger(\tau_{k-1})\,
\Delta \tau$ but then the exponent on the r.h.s. of \eq{18} should be
expanded too. If the derivative is large so that one cannot expand
the exponent (and FITZ keep it), then \eq{18} is simply wrong. To see it
more clearly let us take an example of a zero field, $A=0,\quad
A^\Omega= -i\dot\Omega\,\Omega^\dagger$. Apparently,

\beq
\Omega(\tau_k)\Omega^\dagger(\tau_{k-1})
\neq
\exp \int_{\tau_{k-1}}^{\tau_k} d\tau\, \dot\Omega\Omega^\dagger
\la{neq}\eeq
if the l.h.s is an arbitrary unitary matrix. The path ordering on the
r.h.s. is absolutely necessary to restore the l.h.s.

Even if one accepts the wrong \eq{18} the way FITZ proceed further
is questionable. According to \eq{vern} FITZ need to compute the
character of \eq{18}. Denoting the exponent in \eq{18} by an $su(2)$
matrix $\hat\Phi$ in representation $J$,

\beq
\hat\Phi=\int d\tau\, A^\Omega =\Phi^aT^a
=\int d\tau\,\left[A^\Omega(\tau)\right]^a\,T^a,
\la{phi}\eeq
the character can be written as
\bea
\nonumber
\chi_J[\exp i\hat\Phi]&=&\sum_{m=-J}^J \exp\, im\sqrt{\Phi^a\Phi^a}
=\sum_{m=-J}^J \exp \,im\sqrt{
\int d\tau_1\left[A^\Omega(\tau_1)\right]^a
\int d\tau_2\left[A^\Omega(\tau_2)\right]^a}\\
\la{char2}
&=&\sum_{m=-J}^J \exp \,im\sqrt{
\int\!\!\int d\tau_1d\tau_2\,c_J\,
\Tr\left[A^\Omega(\tau_1)A^\Omega(\tau_2) \right]},\\
\nonumber
c_J&=&\frac{3}{J(J+1)(2J+1)}.
\eea
Meanwhile, FITZ use the following formula (their eqs.(19), (21) and
(22))

\beq
\chi_J[\exp i\hat\Phi]=\sum_{m=-J}^J \exp \,im \int d\tau\,
\sqrt{2\, \Tr\left[A^\Omega(\tau) A^\Omega(\tau)\right]}.
\la{wrong3}\eeq
Leaving aside the incorrect numerical coefficient,
the square root of a product of integrals is {\em not equal} to the
integral of the square root of the product of integrands. In this
way FITZ arrive to their formula for the Wilson loop,
\eq{30}. Their $SU(3)$ generalization (section 3 of \cite{FITZ})
is based on the same manipulation and is wrong from the start.\\

To conclude this section: all points of the FITZ' criticism of our
formula for the Wilson loop are based on their errors in mathematics,
and the alternative formula suggested by these authors is
mathematically inconsistent, too.  \\

Finally, we would like to comment on two sentences from FITZ' paper.
In the Conclusion they write: ``The use of the erroneous path integral
representation for Wilson loops [meaning our formula] has led to the
conclusion that at large distances the average value of the Wilson
loops shows area-law falloff for any irreducible representation of
$SU(N)$...", and ``...has led to result supporting the hypothesis of
maximal Abelian projection". We are not aware of any work where
either of the statements has been mathematically derived using our
formula.

Obviously on the contrary, the non-Abelian Stokes theorem of section 3
stresses the difference between various representations for Wilson
loops: in $SU(2)$ there is a clear distinction between integer and
half-integer representations (see, e.g., \eq{moncontr}); in higher
groups the number of integration variables itself depends on the
representation (see the end of section 3).

\section{Lattice-regularized formula for the Wilson loop}

The Wilson loop on a lattice is just the trace of the product of link
variables in a given representation along the discretized contour.
Else, it is the character of the product of link variables,

\beq
W_J=\frac{1}{2J+1}\chi_J\left(U_{N,N-1}U_{N-1,N-2}\ldots U_{1,N}\right)
\la{Wlat1}\eeq
(we consider the $SU(2)$ gauge group for simplicity).
The aim of this section is to derive a representation for
\eq{Wlat1} analogous to the continuum \eq{W12}, using lattice
regularization. It has the form:

\beq
W_J={\cal N}^{-1}\int \prod_{k=1}^N dS_k\,\exp\, \frac{z}{2}\,
\Tr\,(S^\dagger_kU_{k,k-1}S_{k-1}\tau_3),
\la{Lattice1}
\eeq
where $z$ is a function of $J$ and ${\cal N}$ is a normalization
coefficient, both of them to be determined below. $U_{k,k-1}$ are link
variables which in the continuum limit become $U_{k,k-1}\approx 1+iaA_k$
where $a$ is the lattice spacing and $A_k$ is the component of the
Yang--Mills field along the link at point $k$. In \eq{Lattice1} one
integrates over all gauge transformations $S_k$ at lattice sites $k$
along the loop, with the condition that $S_0=S_N$ where $N$ is the total
number of links in the loop. Integration is over the Haar measure
normalized to unity. We shall see that \eq{Lattice1} is valid for large
Wilson loop with number of links $N\gg 1$. Notice that on cubic lattices
$N$ is always even for closed loops.

In the limit $aA_k\ll 1$ and of smoothly varying $A_k$ and $S_k$
\eq{Lattice1} coincides with \eq{W12} provided one takes $z=2J$. However,
\eq{W12} has been derived from another regularization of the functional
integral over gauge transformations, and there is no {\it a priori}
reason to expect that in the lattice regularization $z$ should be the
same.

Let us expand the exponent in \eq{Lattice1} according to \eq{corr1}:
\bea
\la{dec1}
W_J&=&{\cal N}^{-1}\int \prod_{k=1}^N dS_k
\sum_{j_k}
\tilde{a}_{j_k}(z)\sum_{m_k=-j_k}^{j_k}
e^{i\pi m_k}
D^{j_k}_{m_km_k}
\left(S^\dagger_kU_{k,k-1}S_{k-1}\right),\\
\la{tildea}
\qquad \tilde a_j(z)&=&e^{-i\pi j}\,(2j+1)\,\frac{2}{z}J_{2j+1}(z).
\eea
Every matrix $S_k$ enters twice this expression. The integration
can be performed as follows:
\beq
\int dS_k\;
D^{j_{k+1}}_{m_{k+1}m_{k+1}}\!\left(AS_{k}\right)
D^{j_{k}}_{m_{k}m_{k}}\!(S^\dagger_{k}B)
=\frac{1}{2j_k+1}\delta_{j_kj_{k+1}}\,\delta_{m_km_{k+1}}
D^{j_k}_{m_{k}m_{k}}\!\left(AB\right).
\la{Lattice2}
\eeq
Using \eq{Lattice2} we get:
\bea
\la{Lattice3}
W_J&=&{\cal N}^{-1}\sum_j \left[b_j(z)\right]^N e^{i\pi N(j-m)}
\sum_m D^{j}_{mm}\left(U_{N,N-1}U_{N-1,N-2}\ldots U_{1,N}\right),\\
\la{b}
b_j(z)&=&\frac{2}{z}J_{2j+1}(z)
\eea
The factor $e^{i\pi N(j-m)}=1$ for any $j,m$ since $j\!-\!m$ is an integer
and $N$ is even. Consequently,
\bea
\la{Lattice4}
W_J&=&{\cal N}^{-1}\sum_j \left[b_j(z)\right]^N
\chi_j\left(U_{N,N-1}U_{N-1,N-2}\ldots U_{1,N}\right),\\
\la{N}
{\cal N}&=&\sum_j(2j+1)\left[b_j(z)\right]^N,
\eea
where we have chosen the normalization factor ${\cal N}$
such that $W_J=1$ for unity link variables.

Thus, \eq{Lattice1} is actually a weighted sum of Wilson loops in all
representations $j$. However at $N \gg 1$ the sum in \eq{Lattice4}
is dominated by the term with $j$ maximizing the coefficient
$b_j(z)$. Then only one term survives both in the numerator and
the denominator of \eq{Lattice4} and we obtain:

\beq
W_J\stackrel{{\small N\gg 1}}{\longrightarrow}\frac{1}{2J+1}
\chi_J\left(U_{N,N-1}U_{N-1,N-2}\ldots U_{1,N}\right),
\la{Lattice51}\eeq
as it should be for the Wilson loop in representation $J$.

Let us choose $z(J)$ from the requirement that $b_j(z)$ is
maximal at $j=J$. We find that for $J\leq 2$ one can choose

\beq
z(J)=2(J+1).
\la{z}\eeq
With this value of $z(J)$ the Wilson loop in the needed
representation $J$ is reproduced by \eq{Lattice1} at large number of
links. [We would like to mention on this occasion that one can obtain
\eq{z} also in a continuum formulation where the regularization is
performed by introducing small moments of inertia
$I_{\perp,\parallel}$, as in ref.  \cite{DP1}. In that paper we used
the following limiting procedure:  first $I_\parallel\to 0$, then
$I_\perp\to 0$, and obtained $z(J)=2J$.  Had we chosen the opposite
order of limits we would get \eq{z}.]

In fact, one is not bound to take simple values like given by \eq{z}
but choose the values of $z(J)$ for $J=\frac{1}{2},1,\frac{3}{2}\ldots$
such that $b_j(z(J))$ is {\em maximum maximorum} at $j=J$. The result
of such `fine tuning' is presented in Table 1, where we give the best
value of $z$ for given $J$, the maximum value of $b_{{\rm best}}$ and
the ratio of $b_j(z)$ to $b_{{\rm best}}$ for the most `dangerous'
competitor at $j\neq J$.

\begin{center} {\bf Table 1}\\
\medskip
\begin{tabular}{|c|c|c|c|}
\hline
$J$     & best $z$ & $b_{{\rm best}}$& $b_j(z)/b_{{\rm best}}\;\;(j)$ \\
\hline
$\frac{1}{2}$& 3.25103 & 0.296 & 0.50 $\;\;\;$ (0) \\
\hline
$1$          & 4.36765 & 0.198 & 0.60 (1/2) \\
\hline
$\frac{3}{2}$& 5.46564 & 0.146 & 0.86 $\;\;\;$ (0) \\
\hline
$2$          & 6.55104 & 0.114 & 0.83 (1/2) \\
\hline
$\frac{5}{2}$& 7.62728 & 0.093 & 0.83 $\;\;\;$ (3) \\
\hline
$3$          & 8.69644 & 0.078 & 0.85  (7/2) \\
\hline
\end{tabular}
\end{center}
If, for example, we wish to get the Wilson loop in representation
$J=\frac{5}{2}$, we choose the coefficient $z(5/2)=7.62728$. The largest
contribution to \eq{Lattice4} will be from $j=J=\frac{5}{2}$. The
next largest but parasite contribution will be from $j=3$; however,
the Wilson loop in this representation will be suppressed as $0.83^N$
as compared to the needed $j=\frac{5}{2}$ contribution. If the number
of links in the loop is $N=16$ the suppression factor is 0.051, for
$N=24$ it is 0.011.

The lattice version of the `non-Abelian Stokes theorem' is trivial:
one takes any surface (realized as a collection of plaquettes) inside
a given discretized contour (realized as a chain of links) and writes
in the exponent a sum of the terms
$\Tr\left[\,S^\dagger_k\,U_{k,k-l}\,S_{k-l}\,\tau_3\right]$ for
each link belonging to all plaquettes inside the contour. Since all
internal links are encountered twice in this sum, once going in a
positive and once in a negative direction, all links cancel except
those lying at the border of the surface, that is on the contour
itself. The cancellation of links is due to the fact that
$\Tr(V\tau_3)=-\Tr(V^\dagger \tau_3)$ for a unitary matrix $V$.
The lattice version of the non-Abelian Stokes theorem reads:

\beq
W_J={\cal N}^{-1}\int \left(\prod_{k\in {\rm surf}} dS_k\right)\,
\exp\left\{ \frac{z}{2}\, \sum_{{\rm plaq}\in {\rm surf}}\;
\sum_{{\rm links} \in {\rm plaq}}\;
\Tr\,(S^\dagger_kU_{k,k-l}S_{k-l}\tau_3)\right\}.
\la{Stokeslattice}\eeq
We would like to stress that any Stokes theorem is trivial when written
in discretized form.  At the same time, if we first assemble links
belonging to one plaquette and ascribe to the quantities $S$ and $U$
arguments corresponding to the plaquette centers, we recover, in the
appropriate continuum limit, the non-Abelian Stokes theorem for the
continuum, \eq{W3}.

If one uses \eq{Lattice1} or \eq{Stokeslattice} for the {\em
average} Wilson loop in the lattice formulation of gauge theory there
is no need to perform explicitly integration over gauge transformations
as these are included in the integration over link variables.
Therefore, one can use the following formula for the average Wilson
loop on the lattice:

\beq
\langle W_J\rangle = {\cal N}^{-1} \prod_l \int dU_l\,
\exp\left[{\rm lattice\;action}\right]\;
\exp \left[\frac{z(J)}{2}\sum_{l\in C}\Tr(U_l\tau_3)\right].
\la{lat}\eeq
Actually, this formula has been suggested in ref. \cite{DP3}, however
with the coefficient $z(J)=2J$ arising from another regularization.
This is not too consistent: in lattice calculations one has to use
lattice regularization for the Wilson loop. Therefore, one has either
to put $z(J)=2(J+1)$ (for small $J$) or, better, pick it up from
Table 1.

Finally, we note that the choice of the matrix $\tau_3$ in \eq{lat} is
arbitrary: one can take any rotated matrix as well. This fact can be
used to increase many times the statistics in numerical computation of
the average Wilson loop from \eq{lat}. It may be interesting to study
the dependence of the r.h.s of \eq{lat} on the coefficient $z$ for a
fixed but large loop as it can provide new information on the
difference of integer and half-integer representations in respect to
confinement.

\section{Conclusions}

We have formulated the non-Abelian Stokes theorem for the Wilson loop.
The path-ordering is replaced by an ordinary exponent of a surface
integral, but at the price of an additional functional integration over
all gauge transformations of the Yang--Mills potential, actually, over
a coset depending on the representation in which the Wilson loop is
considered. We have given several forms of this representation, and
discussed requirements on the continuation of the fields inside the
contour.

Since the validity of our formula for the Wilson loop, which is key to
the non-Abelian Stokes theorem, has been questioned recently in
ref. \cite{FITZ} we had to reply to that criticism. We have demonstrated
that it is groundless as due to mistakes in mathematics, which we have
thoroughly pinpointed, one by one.

As the lattice regularization of functional integrals is one of the
most popular we have included the derivation of the formula for the
Wilson loop in lattice regularization. The resulting \eq{Lattice1}
is very similar to the continuum formula \ur{W12}. The corresponding
lattice version of the Stokes theorem is almost trivial.

A formula analogous to the non-Abelian Stokes theorem can be also
derived in gravity theory for holonomies in curved spaces \cite{DP4}.\\

One of us (V.P.) thanks NORDITA for kind hospitality and the
Russian Foundation for Basic Research for partial support, grant
RFBR-00-15-96610.

\end{document}